\documentstyle[aps,preprint,eqsecnum]{revtex}
\begin{document}
\draft
\preprint{IM SB RAS NNA-4-96}
\title{Heavy isovector  resonances in  $e^+e^-$ annihilation, in the decay
$J/\psi\to\pi^+\pi^-\pi^0$ and in the reaction $K^-p\to\pi^+\pi^-\Lambda$.}
\author{N.N.~Achasov \footnote{e-mail: achasov@math.nsk.su}
 and A.A.~Kozhevnikov}
\address{Laboratory of Theoretical Physics,\\
S.L.~Sobolev Institute for Mathematics,\\
630090, Novosibirsk 90, Russian Federation}
\date{\today}
\maketitle
\language=1
\begin{abstract}
The results of an analysis are presented of some recent data on the
reactions $e^+e^-\to\pi^+\pi^-\pi^+\pi^-$,
$e^+e^-\to\pi^+\pi^-\pi^0\pi^0$ with the subtracted $\omega\pi^0$
events, $e^+e^-\to\omega\pi^0$,  $e^+e^-\to\eta \pi^+\pi^-$,
$e^+e^-\to\pi^+\pi^-$, $J/\psi\to\pi^+\pi^-\pi^0$ and
$K^-p\to\pi^+\pi^-\Lambda$, upon taking into account both the strong energy
dependence of the partial widths on energy and the previously neglected
mixing of the $\rho$ type resonances. The above effects are shown to exert
an essential influence on the specific values of masses and coupling
constants of heavy resonances and hence are necessary to be accounted for in
establishing their true nature.
\end{abstract}


\newpage
\section{Introduction}
\label{sec:intro}

          The potential interest in the energy range between 1 and 2 GeV
in the  $e^+e^-$ annihilation is determined by the presence there both the
higher excitations of the ground state vector nonet and possible existence
of exotic non-$q\bar q$- states possessing the quantum numbers allowed in the
$q\bar q$ quark model. Earlier attempts of the interpretation of the
observed resonance structures as manifestations of the production of the
states with hidden exotics \cite{bityuk87} were based on naive pure Breit-
Wigner model of the production amplitude with the supposed fixed partial
widths and with the neglect of the complicated dynamics of  reactions.
In the meantime the above energy range is characterized by the opening of
large number of the multiparticle hadronic decay modes. Their influence on
the dynamics of the amplitudes is caused by the fact that the signals with
the hidden exotics themselves can be imitated by the decays of the higher
excitations of the ground state vector nonet proceeding via two-step
processes with those hadronic states as intermediate ones \cite{ach88}.
The fast growth of the partial widths with energy side by side with their
large magnitudes become essential. This results in an appreciable mixing
via the common decay channels between both the heavy resonances and between
them and the states from the ground state nonet.

        Taking into account the crucial role of heavy  $\rho^\prime$
resonances in the problem of the identification of the states with hidden
exotics, we consistently take into account in the present paper
the above mentioned effects of the mixing and of the fast energy growth of
the partial widths.  To this end the production of such resonances is
analyzed in the $I=1$ channel of the $e^+e^-$ annihilation for the final
states  $\pi^+\pi^-$ \cite{{barkov85},{bisello89}},
$\omega\pi^0$ \cite{{nd},{dm2}},  $\pi^+\pi^-\pi^+\pi^-$
\cite{{nd},{dm2}}, $\pi^+\pi^-\pi^0\pi^0$ with the subtracted 
$\omega\pi^0$ events \cite{{nd},{dm2}},
$\eta \pi^+\pi^-$ \cite{anton88}, in the decay
$J/\psi\to\pi^+\pi^-\pi^0$ \cite{mark} and in the reaction
$K^-p\to\pi^+\pi^-\Lambda$ \cite{lass}.
These effects were ignored in earlier works  \cite{{erkal86},{don}},
devoted to the analysis of the  $e^+e^-$ annihilation data and in the papers
\cite{{mark},{lass}} dealing with the reactions  $J/\psi\to\pi^+\pi^-\pi^0$
and $K^-p\to\pi^+\pi^-\Lambda$.

	The following material is organized as follows. Sec. \ref{sec2}
contains the expressions for the relevant reaction cross section and for the
mass spectrum of the $\pi^+\pi^-$ pair in the decay $J/\psi\to\pi^+\pi^-\pi^0$.
The results of the analysis of some recent data
\cite{{barkov85},{bisello89},{nd},{dm2},{anton88},{mark},{lass}} in the
framework of the approach when the resonances $\rho(770)$, $\rho^{\prime}_1$
, $\rho^{\prime}_2$ and their mixing are taken into account are given in
Sec. \ref{sec3}.  Sec. \ref{sec4} is devoted to the conclusions drawn from the
data analysis.

\section{Expressions for the cross sections and mass spectrum.}
\label{sec2}

	Let us give the expressions for the cross sections of the reactions of
interest taking into account the mixing of the resonances $\rho(770)$,
$\rho^{\prime}_1$ and $\rho^{\prime}_2$ in the frame work of the field theory
inspired approach \cite{ach84} based on the summation of the loop corrections
to the propagators of the unmixed states. The virtue of this approach is
that corresponding amplitudes obey the unitarity requirements. First consider
the final states with the simple reaction dynamics,
$e^+e^-\to\pi^+\pi^-$, $\omega\pi^0$ and  $e^+e^-\to\eta \pi^+\pi^-$,
leaving for a while dynamically more involved final states
$\pi^+\pi^-\pi^+\pi^-$ and $\pi^+\pi^-\pi^0\pi^0$ with subtracted
$\omega\pi^0$ events. One has:
\begin{equation}
\sigma(e^+e^-\to\rho+\rho^{\prime}_1+\rho^{\prime}_2\to f)=
\frac{4\pi\alpha^2}{s^{3/2}}\Biggl|\Biggl(\frac{m^2_\rho}{f_\rho},
\frac{m^2_{\rho^{\prime}_1}}{f_{\rho^{\prime}_1}},
\frac{m^2_{\rho^{\prime}_2}}{f_{\rho^{\prime}_2}}\Biggr)G^{-1}(s)
\pmatrix{g_{\rho f}\cr g_{\rho^{\prime}_1 f}\cr g_{\rho^{\prime}_2 f}
\cr}
\Biggr|^2P_f,
\label{eq1}
\end{equation}
where $f=\pi^+\pi^-$, $\omega\pi^0$ and $\eta \pi^+\pi^-$; $s$ is the total
center-of-mass energy squared, $\alpha=1/137$. For a purpose of uniformity
of the expression Eq. (\ref{eq1}) in the case of the $\pi^+\pi^-$ channel
the contribution of the $\rho\omega$ mixing is omitted for a while. It will
be restored later on. The leptonic widths of the unmixed states are expressed
through the leptonic coupling constants $f_{\rho_i}$ as usual:
\begin{equation}
\Gamma_{\rho_ie^+e^-}=\frac{4\pi\alpha^2}{3f_{\rho_i}^2}m_{\rho_i}.
\label{eq1a}
\end{equation}
The factor $P_f$ for the mentioned final states reads, respectively
\begin{equation}
P_f\equiv P_f(s)=
\frac{2}{3s}q^3_{\pi\pi}\mbox{, }\frac{1}{3}q^3_{\omega\pi}\mbox{, }
\frac{1}{3}\langle q^3_{\rho\eta}\rangle\cdot\frac{2}{3}.
\label{eq2}
\end{equation}
The multiplier  2/3 in the case of  $\eta\pi^+\pi^-$ arising in the simplest
quark model relates the  $\rho\eta$ and $\omega\pi^0$ production amplitudes,
provided the pseudoscalar mixing angle is taken to be $\theta_P=-11^o$, and
\begin{equation}
\langle q^3_{\rho\eta}\rangle=\int\limits_{(2m_\pi)^2}
^{(\sqrt{s}-2m_\eta)^2}dm^2
\rho_{\pi\pi}(m)q^3(\sqrt{s},m,m_\eta).
\label{eq2a}
\end{equation}
Hereafter the function  $\rho_{\pi\pi}(m)$ is aimed to account for the finite
width of the intermediate $\rho(770)$ meson and looks as
\begin{equation}
\rho_{\pi\pi}(m)=\frac{\frac{1}{\pi}m\Gamma_\rho(m)}
{(m^2-m^2_\rho)^2+(m\Gamma_\rho(m))^2},
\label{eq5}
\end{equation}
where $\Gamma_\rho(m)$ is the width of the  $\rho$ meson determined mainly by
the $\pi^+\pi^-$ decay while
\begin{equation}
q_{ij}\equiv q(M,m_i,m_j)=\frac{1}{2M}\sqrt{[M^2-(m_i-m_j)^2]
[M^2-(m_i+m_j)^2]}
\label{eq6}
\end{equation}
is the magnitude of the momentum of either particle $i$ or $j$, in the rest
frame of the decaying particle.

	Let us make some remarks about the way of accounting for of the
$\pi^+\pi^-\pi^+\pi^-$ and $\pi^+\pi^-\pi^0\pi^0$ decay modes.
The details of the mechanisms of the decays
$\rho^{\prime}_{1,2}\to\pi^+\pi^-\pi^+\pi^-$ and
$\rho^{\prime}_{1,2}\to\pi^+\pi^-\pi^0\pi^0$  with subtracted
$\omega\pi^0$ events are still poorly understood. The $4\pi$ mode is known
to originate from the  $\rho\pi\pi$ states. Guided by the isotopic invariance,
one can express the amplitudes of production of the specific charge
combinations through the amplitudes $M_I$ with the given isospin $I$
of the final pion pair. Then the relation between the I=0 and
I=2 amplitudes results from the absence of
$\rho^0\pi^0\pi^0$ \cite{{nd},{dm2}} which, in turn, permits one to write
\begin{eqnarray}
M(\rho^\prime\to\rho^0\pi^+\pi^-)&=&M_2,     \nonumber\\
M(\rho^\prime\to\rho^\mp\pi^\pm\pi^0)&=&{1\over2}(M_2+M_1).
\label{isot}
\end{eqnarray}
The amplitude $M_2$ will be further taken into account as a pointlike vertex
$\rho^{\prime}_{1,2}\to\rho^0\pi^+\pi^-$. Such an approximation
seems to be justifiable since the possible $a_1(1260)\pi$ and $h_1(1170)\pi$
intermediate states containing the axial vector mesons have two partial
waves in their decay into $\rho\pi$, thus resulting in a structureless
angular distribution of final pions. An analogous form is assumed
for the vertex $\rho^0(770)\to\rho^0(770)\pi^+\pi^-$ for the s-channel $\rho$
meson lying off its mass shell. Taking into account the
vector current conservation, the relation
$g_{\rho^0\rho^0\pi^+\pi^-}=2g^2_{\rho\pi\pi}$ can be considered as
a guide for corresponding coupling constant \cite{fn1}.
The amplitude $M_1$ corresponds to the decay $\rho^\prime\to\rho^+\rho^-$.
Having in mind all these remarks, one can write the expressions for the
cross sections. One has
\begin{equation}
\sigma(e^+e^-\to\pi^+\pi^-\pi^+\pi^-)=
\frac{(4\pi\alpha)^2}{s^{3/2}}\Biggl|\Biggl(\frac{m^2_\rho}{f_\rho},
\frac{m^2_{\rho^{\prime}_1}}{f_{\rho^{\prime}_1}},
\frac{m^2_{\rho^{\prime}_2}}{f_{\rho^{\prime}_2}}\Biggr)G^{-1}(s)
\pmatrix{2g^2_{\rho\pi\pi}\cr g_{\rho^{\prime}_1\rho^0\pi^+\pi^-}
\cr g_{\rho^{\prime}_2\rho^0\pi^+\pi^-}\cr}
\Biggr|^2W_{\pi^+\pi^-\pi^+\pi^-}(s)
\label{4pic}
\end{equation}
in the case of the final state $\pi^+\pi^-\pi^+\pi^-$ and
\begin{eqnarray}
\sigma(e^+e^-\to\pi^+\pi^-\pi^0\pi^0)&=&
\frac{(4\pi\alpha)^2}{s^{3/2}}\Biggl\{{1\over2}
\Biggl|\Biggl(\frac{m^2_\rho}{f_\rho},
\frac{m^2_{\rho^{\prime}_1}}{f_{\rho^{\prime}_1}},
\frac{m^2_{\rho^{\prime}_2}}{f_{\rho^{\prime}_2}}\Biggr)G^{-1}(s)
\pmatrix{2g^2_{\rho\pi\pi}\cr g_{\rho^{\prime}_1\rho^0\pi^+\pi^-}
\cr g_{\rho^{\prime}_2\rho^0\pi^+\pi^-}\cr}
\Biggr|^2W_{\pi^+\pi^-\pi^+\pi^-}(s)+     \nonumber\\
& &\Biggl|\Biggl(\frac{m^2_\rho}{f_\rho},
\frac{m^2_{\rho^{\prime}_1}}{f_{\rho^{\prime}_1}},
\frac{m^2_{\rho^{\prime}_2}}{f_{\rho^{\prime}_2}}\Biggr)G^{-1}(s)
\pmatrix{g_{\rho\pi\pi}\cr g_{\rho^{\prime}_1\rho^+\rho^-}
\cr g_{\rho^{\prime}_2\rho^+\rho^-}\cr}
\Biggr|^2W_{\pi^+\pi^-\pi^0\pi^0}(s)\Biggr\}
\label{4pin}
\end{eqnarray}
in the case of the final state $\pi^+\pi^-\pi^0\pi^0$ with subtracted
$\omega\pi^0$ events. Note that in view of universality of the  $\rho$
coupling, the relation  $g_{\rho^0\rho^+\rho^-}=g_{\rho^0\pi^+\pi^-}$ holds.
The final state factors in the above expressions are, respectively,
\begin{eqnarray}
W_{\pi^+\pi^-\pi^+\pi^-}(s)&=&\frac{1}{(2\pi)^34s}\int
\limits_{(2m_\pi)^2}^{(\sqrt{s}-2m_\pi)^2}dm^2_1
\rho_{\pi\pi}(m_1)\int\limits_{(2m_\pi)^2}^{(\sqrt{s}-m_1)^2}dm^2_2
\cdot           \nonumber\\
& &\biggl(1+
\frac{q^2(\sqrt{s},m_1,m_2)}{3m^2_1}\biggr)
q(\sqrt{s},m_1,m_2)q(m_2,m_\pi,m_\pi),   \nonumber\\
W_{\pi^+\pi^-\pi^0\pi^0}(s)&=&\frac{1}{2\pi s}\int
\limits_{(2m_\pi)^2}^{(\sqrt{s}-2m_\pi)^2}dm^2_1
\rho_{\pi\pi}(m_1)\int\limits_{(2m_\pi)^2}^{(\sqrt{s}-m_1)^2}dm^2_2
\rho_{\pi\pi}(m_2)q^3(\sqrt{s},m_1,m_1).
\label{phsp}
\end{eqnarray}

	The matrix of inverse propagators looks as
\begin{equation}
G(s)=\pmatrix{D_\rho&-\Pi_{\rho\rho^{\prime}_1}&-\Pi_{\rho\rho^{\prime}_2}\cr
-\Pi_{\rho\rho^{\prime}_1}&D_{\rho^{\prime}_1}&-\Pi_{\rho^{\prime}_1
\rho^{\prime}_2}\cr
-\Pi_{\rho\rho^{\prime}_2}&-\Pi_{\rho^{\prime}_1\rho^{\prime}_2}&
D_{\rho^{\prime}_2}\cr}.
\label{eq7}
\end{equation}
It contains the inverse propagators of the unmixed states
$\rho_i=\rho(770)\mbox{, }
\rho^{\prime}_1\mbox{, }\rho^{\prime}_2$,
\begin{equation}
D_{\rho_i}\equiv D_{\rho_i}(s)=m^2_{\rho_i}-s-i\sqrt{s}\Gamma_{\rho_i}(s),
\label{eq8}
\end{equation}
where
\begin{eqnarray}
\Gamma_{\rho_i}(s)&=&\frac{g^2_{\rho_i\pi\pi}}{6\pi s}q^3_{\pi\pi}+
\frac{g^2_{\rho_i\omega\pi}}{12\pi}\biggl(q^3_{\omega\pi}+q^3_{K^{\ast}K}+
\frac{2}{3}\langle q^3_{\rho\eta}\rangle\biggr)+{3\over2}
g^2_{\rho_i\rho^0\pi^+\pi^-}W_{\pi^+\pi^-\pi^+\pi^-}(s)+    \nonumber\\
& &g^2_{\rho_i\rho^+\rho^-}W_{\pi^+\pi^-\pi^0\pi^0}(s)
\label{eq9}
\end{eqnarray}
are the energy dependent widths, and the nondiagonal polarization operators
$$\Pi_{\rho_i\rho_j}=\mbox{Re}\Pi_{\rho_i\rho_j}+i\mbox{Im}
\Pi_{\rho_i\rho_j}$$
describing the mixing. Their real part is still unknown and further will be
assumed to be some constant while the imaginary part is given by the unitarity
relation as
\begin{eqnarray}
\mbox{Im}\Pi_{\rho_i\rho_j}&=&\sqrt{s}
\biggl[\frac{g_{\rho_i\pi\pi}g_{\rho_j\pi\pi}}{6\pi s}q^3_{\pi\pi}+
\frac{g_{\rho_i\omega\pi}g_{\rho_j\omega\pi}}{12\pi}
\biggl(q^3_{\omega\pi}+q^3_{K^{\ast}K}+
\frac{2}{3}\langle q^3_{\rho\eta}\rangle\biggr)+     \nonumber\\
& & {3\over2}g_{\rho_i\rho^0\pi^+\pi^-}g_{\rho_j\rho^0\pi^+\pi^-}
W_{\pi^+\pi^-\pi^+\pi^-}(s)+
g_{\rho_i\rho^+\rho^-}g_{\rho_j\rho^+\rho^-}
W_{\pi^+\pi^-\pi^0\pi^0}(s)\biggr].
\label{eq10}
\end{eqnarray}
The states vector+pseudoscalar in Eqs. (\ref{eq9}) and (\ref{eq10}) are taken
into account assuming the quark model relations between their couplings with
the $\rho_i$ resonances. However, the possibility of the violation of these
relations will be included below. The $K\bar K$ final states in the decays
of the  $\rho^{\prime}_{1,2}$ have relatively small branching ratios
\cite{pdg} and hence are neglected. In the meantime the decay
$\rho(770)\to K\bar K$ at $\sqrt{s}>1$ GeV is included assuming the quark
model relation for the off-mass-shell $\rho(770)$ couplings.

To describe the MARK III data \cite{mark} on the $\pi^+\pi^-$ mass spectrum
in the decay $J/\psi\to\pi^+\pi^-\pi^0$, taking into account the cut in the
cosine of the angle of the momentum of outgoing pions in the $\pi^+\pi^-$
rest system relative to the $\rho^0$ momentum in the c.m.s.,
$|\cos\theta_1|\leq 0.2$, one can use the expression
\begin{eqnarray}
{d\Gamma\over dm}(J/\psi\to\pi^+\pi^-\pi^0)&=&{1\over6(2\pi)^3}
[q(m_{J/\psi},m,m_\pi)q(m,m_\pi,m_\pi)]^3
\int\limits_{-0.2}^{0.2}d\cos\theta_1\sin^2\theta_1\cdot  \nonumber\\
& &|A(m^2)+A(m^2_+)+A(m^2_-)|^2
\label{spec}
\end{eqnarray}
with
\begin{equation}
A(m^2)=(F_1,F_2,F_3)G^{-1}(m^2)
\pmatrix{g_{\rho\pi\pi}\cr g_{\rho^{\prime}_1\pi\pi}
\cr g_{\rho^{\prime}_2\pi\pi}\cr},
\label{spec1}
\end{equation}
where
\begin{equation}
m_\pm^2={1\over2}(m_{J/\psi}^2+3m^2_\pi-m^2)\pm2 m_{J/\psi}
q(m_{J/\psi},m,m_\pi)
q(m,m_\pi,m_\pi)\cos\theta_1/m,
\label{mplu}
\end{equation}
$F_{1,2,3}$ are the  production amplitudes of the $\rho(770)$,
$\rho^\prime_1$ and $\rho^\prime_2$ resonances in the $J/\psi$ decays whose
explicit form will be specified below, $m$ being an invariant mass of the
$\pi^+\pi^-$ pair.

To fit the LASS data \cite{lass} on the modulus of the $\pi^+\pi^-$
production amplitude in the reaction $K^-p\to\pi^+\pi^-\Lambda$  one can use
the expression analogous to Eq. (\ref{eq1}) in the case of $f=\pi^+\pi^-$,
but without the factor  $s^{-3/2}$ pertinent to the one-photon
$e^+e^-$ annihilation and with the proper relative production amplitudes
instead of $m^2_{\rho_i}/f_{\rho_i}$ as is exemplified in Eq. (\ref{spec1}).

\section{Results and discussion.}
\label{sec3}

	Let us describe briefly the procedure of the fit to the data. Since the
data on the reactions under consideration,
\begin{equation}
e^+e^-\to\pi^+\pi^-\pi^+\pi^-,
\label{31c}
\end{equation}
\begin{equation}
e^+e^-\to\pi^+\pi^-\pi^0\pi^0
\label{31d}
\end{equation}
with subtracted $\omega\pi^o$ events,
\begin{equation}
e^+e^-\to\omega\pi^0,
\label{31b}
\end{equation}
\begin{equation}
e^+e^-\to\pi^+\pi^-\eta,
\label{31e}
\end{equation}
\begin{equation}
e^+e^-\to\pi^+\pi^-,
\label{31a}
\end{equation}
\begin{equation}
J/\psi\to\pi^+\pi^-\pi^0,
\label{31f}
\end{equation}
\begin{equation}
K^-p\to\pi^+\pi^-\Lambda
\label{31g}
\end{equation}
are gathered in diverse experiments and still possess large uncertainties,
the fit is carried out for each channel separately by means of the $\chi^2$
minimization. The fitted parameters  for the reactions in  $e^+e^-$
annihilation are
\begin{equation}
m_{\rho^{\prime}_{1,2}}\mbox{, }g_{\rho^{\prime}_{1,2}\pi^+\pi^-}\mbox{, }
g_{\rho^{\prime}_{1,2}\omega\pi}\mbox{, }
g_{\rho^{\prime}_{1,2}\rho^0\pi^+\pi^-}\mbox{, }
g_{\rho^{\prime}_{1,2}\rho^+\rho^-}\mbox{, }
f_{\rho^{\prime}_{1,2}}\mbox{, Re}\Pi_{\rho^{\prime}_1\rho^{\prime}_2}.
\label{eq32}
\end{equation}
while for the reactions (\ref{31f}) and (\ref{31g}) one should take the
relative production
amplitudes $F_1$, $F_2$ and $F_3$ [see Eq. (\ref{spec})] instead of the
leptonic couplings, because the production mechanisms in these reactions are
different. The real parts of nondiagonal polarization operators
$\mbox{Re}\Pi_{\rho\rho^{\prime}_1}$ and
$\mbox{Re}\Pi_{\rho\rho^{\prime}_2}$ are set to zero. Indeed, in a sharp
distinction with the imaginary parts Im$\Pi_{\rho_i\rho_j}$ fixed by the
unitarity relation, the real parts  Re$\Pi_{\rho_i\rho_j}$ cannot be evaluated
at present and should be treated as free parameters. However, one must have
in mind that nonzero $\mbox{Re}\Pi_{\rho\rho^{\prime}_{1,2}}\not=0$
result in an appreciable mass shift of the $\rho(770)$ resonance,
\begin{equation}
\delta m_\rho\simeq-\frac{1}{2m_\rho}\mbox{Re}
\biggl[\frac{\Pi^2_{\rho\rho^{\prime}_{1,2}}(m_\rho)}
{m^2_{\rho^{\prime}_{1,2}}-m^2_\rho-im_\rho
(\Gamma_{\rho^{\prime}_{1,2}}(m_\rho)
-\Gamma_\rho(m_\rho))}\biggr],
\label{eq33}
\end{equation}
\cite{{ach92},{fn2}}.
Since the minimization fixes only the combination $m_\rho+\delta m_\rho$,
it is natural to assume that the dominant contribution to the mass
renormalization Eq.(\ref{eq33}) coming from
$(\mbox{Re}\Pi_{\rho\rho^{\prime}_{1,2}})^2$ is already subtracted, so that
the mass of the  $\rho$ meson minimizing the $\chi^2$ function differs from
the actual position of the $\rho$ peak by the magnitudes quadratic in
$\mbox{Im}\Pi_{\rho\rho^{\prime}_{1,2}}$. In practice it manifests itself
in that one seeks the minimum of the  $\chi^2$ given by the values of
$m_\rho$
which are near 770 MeV. Then the minimization procedure automatically
chooses (with rather large errors, of course) the values of
$\mbox{Re}\Pi_{\rho\rho^{\prime}_{1,2}}$ lying close to zero.
By this reason, heaving in mind the existing accuracy of the data, it is
natural to set these parameters to zero from the very start. These
considerations justifiable in the case of the  $\rho(770)$ resonance whose
$q\bar q$ nature is firmly established, however, cannot be applied to the
case of heavy $\rho^\prime_{1,2}$ resonances. The nature of these resonances
is in fact not yet established, and one cannot exclude that they may contain
an appreciable portion of exotics like $q\bar qg$, $q\bar qq\bar q$ etc.
\cite{kalash}.
Hence it is a matter of principle to extract from the existing data the values
of masses and coupling constants of bare unmixed states in order to compare
them with current predictions. By this reason
$\mbox{Re}\Pi_{\rho^\prime_1\rho^\prime_2}$ is considered to be a free
parameter.

	The parameters of the  $\rho(770)$ are chosen from the fitting to the pion
formfactor from the threshold to 1 GeV upon taking into account both the
$\rho\omega$ mixing and the mixing of the $\rho(770)$ with the
$\rho^\prime_{1,2}$ resonances originating from their common decay modes.
To allow for the  $\rho\omega$ mixing, one should add the term
$$\frac{m^2_\rho m^2_\omega\Pi_{\rho\omega}}
{f_\rho f_\omega D_\rho D_\omega},$$
to the expression in between the modulus sign in Eq. (\ref{eq1}) in the case
of the $\pi^+\pi^-$ channel, where
\begin{eqnarray}
D_\omega&\equiv& D_\omega(s)=
m^2_\omega-s-i\sqrt{s}\Gamma_\omega(s),  \nonumber\\
\Gamma_\omega(s)&=&\frac{g^2_{\omega\rho\pi}}{4\pi}W_{3\pi}(s)
\label{dom}
\end{eqnarray}
are respectively the inverse propagator  of the $\omega$ meson and its width
determined mainly by the  $\pi^+\pi^-\pi^0$ decay mode while
$W_{3\pi}(s)$ stands for the phase space volume of the final  $3\pi$ state
(see its expression in e.g.  \cite{ach92}).
Here the real part of the polarization operator of the  $\rho\omega$
transition is taken in the form
\begin{equation}
\mbox{Re}\Pi_{\rho\omega}=2m_\omega\delta_{\rho\omega}+
\frac{4\pi\alpha m^2_\rho m^2_\omega}{f_\rho f_\omega}(1/m^2_\omega-1/s);
\label{eq34}
\end{equation}
$\delta_{\rho\omega}$ is the amplitude of the  $\rho\omega$ transition as
measured at the $\omega$ mass while the last term is aimed to take into
account the fast varying one photon contribution. The expression for imaginary
part $\mbox{Im}\Pi_{\rho\omega}$ is given in  \cite{ach92a}. Note that
$g_{\rho\omega\pi}=g_{\omega\rho\pi}=14.3\mbox{ GeV}^{-1}$ \cite{nd}.
The $\rho(770)$ parameters obtained from fitting the
$e^+e^-\to\pi^+\pi^-$ channel are: $m_\rho=774\pm10\mbox{ MeV, }
g_{\rho\pi\pi}=5.9\pm0.2\mbox{, }
f_{\rho}=5.1\pm0.2\mbox{ ¨ }\delta_{\rho\omega}=2.4\pm1.4$ MeV. The error bars
are determined from the function  $\chi^2$. Notice that the $\rho\omega$
mixing in the reaction  (\ref{31f}) is neglected, since the MARK III data
show no $\rho\omega$ interference pattern and hence the fit is insensitive
to  additional free parameters characterizing the $\omega\pi$ coupling
to the $J/\psi$. As for the reaction
(\ref{31g}), the amplitude of the  $\rho\omega$ transition
$\delta_{\rho\omega}$ is fixed to be  2.4 MeV, while the relative
$\omega$ production amplitude is varied. The fit of the LASS data \cite{lass}
turns out to be insensitive to the specific value of that amplitude.

	It seems to be rather natural that the function of 13 variables
Eq. (\ref{eq32}), $\chi^2$,  possesses a number of local minima characterized
by the parameters considerably differing (by more than 3 standard deviations)
from channel to channel, Eq. (\ref{31c})-(\ref{31g}). The final choice is
implemented under the demand of possibility of the simultaneous fit to all
the channels in the framework of an uniform approach. The results are
presented in Table \ref{tab1} and in Figs. \ref{fig1}-\ref{fig7}.
Taking into account yet  large uncertainties of the data, one can
conclude that the magnitudes of the parameters obtained from the fitting of
diverse channels do not contradict each other.

Let us dwell on the role played by the coupling constant
$g_{\rho^\prime_{1,2}\rho^+\rho^-}$ in the analysis of the reaction
Eq. (\ref{31d}) with the subtracted $\omega\pi^0$ events.
The current data on this channel are contradictory (see Fig.
\ref{fig2}).
In the meantime the ND \cite{nd} and DM2 \cite{dm2} data on the reaction
$e^+e^-\to\pi^+\pi^-\pi^+\pi^-$, Fig. {\ref{fig1}, are consistent in the
 region of overlap. So it would be quite natural to take the parameters of
resonances  extracted from fitting the reaction Eq. (\ref{31c}) and to vary
them within the error bars in order to describe both the data
\cite{nd} with $\sqrt{s}\leq1.4$ GeV and the nonoverlapping data
\cite{dm2} with   $\sqrt{s}>1.7$ GeV. As appears, the better description is
achieved under introduction of a nonzero $g_{\rho^\prime_1\rho^+\rho^-}=7\pm3$
in the case of the channel $e^+e^-\to\pi^+\pi^-\pi^0\pi^0$, in the meantime
the analysis of the remaining channels gives the magnitudes of this coupling
constant  which do not contradict zero (see the Table \ref{tab1})
\cite{atkin85}.
If the value of $g_{\rho^\prime_{1,2}\rho^+\rho^-}\equiv0$ were fixed from
the very start, the central value of the leptonic width
$\Gamma_{\rho^\prime_1ee}=12.9^{+4.5}_{-4.3}$ keV extracted from
the reaction Eq. (\ref{31d}) would deviate by more than the factor of two
from the value $\Gamma_{\rho^\prime_1ee}=5.2^{+2.2}_{-1.9}$ keV obtained from
the reaction Eq. (\ref{31c}), though not going from the double standard
deviation. The final resolution could be possible only after gathering
new consistent data.

The bare 
masses of the resonances  $\rho^\prime_{1,2}$ are seen to be considerably
higher than the actual position of the peaks or structures in cross sections
and mass spectra. This is explained by the two reasons. First, the fast
growth of the partial widths with energy results in the shift
\begin{equation}
\delta   m_{\rho^{\prime}_{1,2}}\sim -\Gamma(s){d\Gamma\over d\sqrt{s}}
(\sqrt{s}=m_{\rho^\prime_{1,2}})
\label{eq35}
\end{equation}
towards the lower values from the bare masses.
Second, there exists the shift due
to the mixing  \cite{ach92} of the upper of two states,
\begin{equation}
\delta m_{\rho^{\prime}_{1,2}}\simeq
\frac{1}{2m_{\rho^{\prime}_{1,2}}}\mbox{Re}
\biggl[\frac{\Pi^2_{\rho\rho^{\prime}_{1,2}}(m_{\rho^{\prime}_{1,2}})}
{m^2_{\rho^{\prime}_{1,2}}-m^2_\rho-im_{\rho^{\prime}_{1,2}}
(\Gamma_{\rho^{\prime}_{1,2}}(m_{\rho^{\prime}_{1,2}})
-\Gamma_\rho(m_{\rho^{\prime}_{1,2}}))}\biggr],
\label{eq36}
\end{equation}
which is negative, since the nondiagonal polarization operator of the
$\rho\rho^{\prime}_{1,2}$ mixing dominates over the
$\rho^{\prime}_1\rho^{\prime}_2$ mixing and possesses in the present case
the imaginary part much greater than the real part. In the meantime the
corresponding mass shift of the  $\rho(770)$ in the $\pi^+\pi^-$ channel due
to the mixing with higher states, see Eq. (\ref{eq33})
turns out to be rather small, $\delta m_\rho\simeq 4$ MeV, since
Im$\Pi_{\rho\rho^{\prime}_{1,2}}$ is small at $\sqrt{s}\simeq m_\rho$.
Plotting the contributions of the unmixed states in Figs. {\ref{fig1}-
\ref{fig4} shows that the dominant contribution to the mass shifts comes
from the fast growth of the partial widths, because the resonance peaks
without the mixing being taken into account are also displaced considerably.

The magnitude of the real part Re$\Pi_{\rho^\prime_1\rho^\prime_2}$
does not contradict zero. However, $\chi^2$ is minimized with nonzero values
of this parameter though with large errors. This means that the quality of the
data is still insufficient to establish
Re$\Pi_{\rho^\prime_1\rho^\prime_2}\not=0$. Note also that a better
description of the channel $\pi^+\pi^-\eta$ is achieved upon introducing
the suppression factor $y_\eta=0.7\pm0.2$ of the coupling constant
$g_{\rho^\prime_1\rho\eta}$ as compared to the value
$\sqrt{2/3}g_{\rho^\prime_1\omega\pi}$ given by the simplest quark model.

The fit to the $\pi^+\pi^-$ mass spectrum in the decay
$J/\psi\to\pi^+\pi^-\pi^0$ with the relative production amplitudes
independent of the $\pi^+\pi^-$ invariant mass gives all the resonance
parameters but $g_{\rho^\prime_1\pi^+\pi^-}=-2.8^{+0.5}_{-0.4}$  
in agreement with other channels.
This latter value  strongly deviates
from $-0.9^{+1.0}_{-1.1}$ and $-1.0\pm0.3$ extracted from the channels
Eq. (\ref{31c}) and
Eq. (\ref{31a}), respectively. Hence, we tried to include the mass dependence
in the simplest linear form $F_i(m^2)=a_i+b_i(m^2-m^2_i)$.
This is equivalent to
the introduction of a background which generates the structure at $\sqrt{s}
\sim1.3$ GeV. As a result, the agreement with other channels is achieved
but at the expence of the poor determination of  the 
$\rho^\prime_1$ mass.
It is this variant that is included in Table \ref{tab1}. Note that the slopes
of the mass dependence $b_{\rho^\prime_{1,2}}$ are compatible with zero
while $b_{\rho(770)}$ is not. Specifically, the relative $\rho(770)$
production amplitude normalized to unity at $m=m_\rho$ is 
$F_\rho(m^2)=1+20^{+3}_{-5}\cdot 10^{-2}\mbox{GeV}^{-2}\times(m^2-m^2_\rho)$.

Note that the Blatt-Weiskopf range 
parameters which are sometimes introduced into the expressions for the partial 
widths of the $\rho(770)$ 
to make acceptable their fast growth with energy, are chosen to be 
zero by the $\chi^2$ minimization. Hence corresponding factors are 
omitted in the expressions for the partial widths. In the meantime, 
the inclusion of such
factors in the case of heavy $\rho^\prime_{1,2}$ resonances is unnecessary
because the energies of the present interest are in the mass range of these
states.

\section{Conclusion}
\label{sec4}

The main conclusion from the present analysis is that the inclusion of both
the mixing of heavy isovector resonances and the energy dependence of their
partial widths is completely necessary when describing the data on the
reactions Eq. (\ref{31c})-(\ref{31g}). The possibility of simultaneous
fit of the existing data and the specific magnitudes of the extracted
parameters necessary for the comparison with current models, are crucially
affected by  these effects. It should be emphasized that the fit to  the
data on the reactions Eq. (\ref{31c}) - (\ref{31e}) alone does not at all
demand the presence of the resonance  $\rho_1^{\prime}$. The large magnitudes
of the coupling constant $g_{\rho_1^{\prime}\omega\pi}$ for these reactions
given in Table \ref{tab1} are already pointed out to be chosen under the
demand of the possibility of simultaneous description of all variety of
data including the reactions (\ref{31a}) - (\ref{31g}). If one discards these
latter reactions, the variants of the fits exist giving
the couplings of the $\rho_1^\prime$ resonance which do not contradict zero.

The novel dynamical feature revealed in the present analysis is the possible
nonzero magnitude of the coupling constants 
$g_{\rho^\prime_{1,2}\rho^+\rho^-}$. The
threshold region in the reaction $e^+e^-\to\pi^+\pi^-\pi^0\pi^0$ with the
subtracted $\omega\pi^0$ events is especially promising for improvement
of the quality of extraction of above coupling constants. The real part of the
polarization operator Re$\Pi_{\rho^\prime_1\rho^\prime_2}$ is  determined
poorly from the current data.  
The study of the  $e^+e^-$ annihilation channels Eq. (\ref{31c}) -
(\ref{31a}) with good statistics and on the same facility is urgent for 
measuring these important dynamical
parameters required for establishing the nature of heavy isovector resonances
and for the final elucidation of the situation.

The present work was supported in part by the grants of Russian Foundation
for Basic Research RFFI-94-02-05, RFFI-96-02-00 and by the grant
INTAS-94-3986.

\begin{figure}
\caption{The result of the description of the reaction
Eq. (\protect\ref{31c}). The data are: ND \protect\cite{nd}, 
DM2 \protect\cite{dm2},
CMD \protect\cite{cmd} and OLYA \protect\cite{olya}. \label{fig1}}
\end{figure}

\begin{figure}
\caption{The result of the description of the reaction
Eq. (\protect\ref{31d}). The data are: ND \protect\cite{nd}, 
DM2 \protect\cite{dm2}, $\gamma\gamma2$ and M3N 
\protect\cite{fran}.    \label{fig2}}
\end{figure}

\begin{figure}
\caption{The result of the description of the reaction
Eq. (\protect\ref{31b}). The data are: Neutral Detector \protect\cite{nd}, 
DM2 \protect\cite{dm2}. \label{fig3}}
\end{figure}

\begin{figure}
\caption{The result of the description of the reaction
Eq. (\protect\ref{31e}). The data are \protect\cite{anton88}. \label{fig4}}
\end{figure}

\begin{figure}
\caption{The result of the description of the reaction
Eq. (\protect\ref{31a}). The data are: OLYA and CMD \protect\cite{barkov85},
DM2 \protect\cite{bisello89}. \label{fig5}}
\end{figure}

\begin{figure}
\caption{The result of the description of the $\pi^+\pi^-$ mass spectrum
\protect\cite{mark} in the decay Eq. (\protect\ref{31f}).  \label{fig6}}
\end{figure}

\begin{figure}
\caption{The result of the description of the data \protect\cite{lass}
on the modulus of the p-wave amplitude
of the reaction Eq. (\protect\ref{31g}).  \label{fig7}}
\end{figure}

\begin{table}
\caption{The magnitudes of the parameters giving the best description
of the data on the reaction Eq. (\protect\ref{31c}) - (\protect\ref{31g}).
The error bars are determined from the function  $\chi^2$. The parameter
$F_{\rho^\prime_{1,2}}=F_{2,3}/F_1$ is the relative production amplitude
 of the corresponding heavy resonance. 
 To avoid the introduction of additional free
 parameters in the case of the reaction Eq. (\protect\ref{31f}), 
 $\mbox{Re}\Pi_{\rho^{\prime}_1\rho^{\prime}_2}$ is fixed to zero,
 while the slope of the mass dependence of the relative production
 amplitude (see the body of the paper) is varied.}
\label{tab1}
\begin{tabular}{cccccccc}
&(\ref{31c})&(\ref{31d})&(\ref{31b})&(\ref{31e})&(\ref{31a})&(\ref{31f})&
(\ref{31g}) \\
\tableline
$m_{\rho^{\prime}_1}\mbox{, GeV}$&$1.35\pm0.05$
&$1.40^{+0.22}_{-0.14}$&$\sim1.4$&$1.46^{+0.30}_{-0.40}$&$1.37^{+0.09}
_{-0.07}$&$1.57^{+0.25}_{-0.19}$&$1.36^{+0.18}_{-0.16}$ \\
$g_{\rho^{\prime}_1\pi^+\pi^-}$&$-0.9^{+1.0}_{-1.1}$&$<18$&$<72\cdot10^{-1}$
&$<39$&$-1.0\pm0.3$&$(-17^{+12}_{-13})\cdot10^{-1}$&$<57\cdot10^{-1}$  \\
$g_{\rho^{\prime}_1\omega\pi}\mbox{, GeV}^{-1}$&$14.9^{+3.6}_{-2.6}$
&$16.6^{+4.6}_{-3.2}$&$<10$&$19^{+11}_{-6}$&$16.6^{+2.2}_{-1.5}$
&$21^{+3}_{-7}$&$13.7^{+4.3}_{-3.2}$\\
$g_{\rho^{\prime}_1\rho^o\pi^+\pi^-}$&$<25$&$<210$&$<210$&$<840$
&$<150$&$<660$&$<540$\\
$g_{\rho^\prime_1\rho^+\rho^-}$&$<72$&$7.1\pm3.0$&undetermined&$<114$&$<45$
&$<57$&$<48$  \\
$F_{\rho^{\prime}_1}$&$2.1^{+0.5}_{-0.4}$&$2.4^{+0.6}_{-0.5}$
&$<66\cdot10^{-1}$&$3.7\pm1.0$&$2.3\pm0.2$&$(12^{+8}_{-2})\cdot10^{-1}$
&$2.1\pm0.5$\\
$m_{\rho^{\prime}_2}\mbox{, GeV}$&$1.851^{+0.027}_{-0.024}$
&$1.79^{+0.11}_{-0.07}$&$1.71\pm0.09$&$1.91^{+1.00}_{-0.37}$
&$1.90^{+0.17}_{-0.13}$&$2.08^{+0.16}_{-0.30}$&-\\
$g_{\rho^{\prime}_2\pi^+\pi^-}$&$1.8\pm1.1$&$<20$&$<33\cdot10^{-1}$
&$<42$&$<18\cdot10^{-1}$&$<33\cdot10^{-1}$&-\\
$g_{\rho^{\prime}_2\omega\pi}\mbox{, GeV}^{-1}$&$-6.1^{+0.7}_{-0.8}$
&$-10\pm3$&$-6.0\pm1.2$&$<45$&$<18$&$-12^{+7}_{-4}$&-\\
$g_{\rho^{\prime}_2\rho^o\pi^+\pi^-}$&$-222\pm9$&$-184^{+23}_{-32}$&
$-188\pm22$&$<960$&$-63^{+19}_{-55}$&$-180^{+190}_{-130}$&-\\
$g_{\rho^\prime_2\rho^+\rho^-}$&$<24$&$<25$&$<48$&$<165$&$<30$&$<114$&-\\
$F_{\rho^{\prime}_2}$&$2.9\pm0.1$&$2.9\pm0.4$&$2.9\pm0.4$&$<2.8$
&$2.0\pm0.8$&$(-18^{+2}_{-6})\cdot10^{-1}$&-\\
$\mbox{Re}\Pi_{\rho^{\prime}_1\rho^{\prime}_2}\mbox{, GeV}^2$
&$<3\cdot10^{-1}$&$<5\cdot10^{-1}$&$<12\cdot10^{-1}$&$<36\cdot10^{-1}$
&$<6\cdot10^{-1}$&$\equiv 0$&-\\
\end{tabular}
\end{table}


\begin{references}
\bibitem{fn1}
The bremstrahlung diagrams with the emission of the  $\rho$ meson
are neglected because of the intermediate pion  is far from mass shell,
so their contribution grows with energy much more slowly as compared to the
pointlike vertex. In any case, taking them into account seems to be premature
since, as is evident from Fig. \ref{fig1},
the contribution of the right  shoulder
of the $\rho(770)$ peak in the  $\pi^+\pi^-\pi^+\pi^-$ channel is much lower
then the contribution of the $\rho^{\prime}_{1,2}$ resonances.
\bibitem{fn2}
This effect is completely analogous to the well known quantum mechanical
repulsion of two levels under the action of perturbation.
\end{references}
\end{document}